\documentclass{article}
\usepackage{graphicx}
\usepackage{amssymb,amsmath}
\def\lsim{\mathrel{\mathchoice {\vcenter{\offinterlineskip\halign{\hfil
$\displaystyle##$\hfil\cr<\cr\sim\cr}}}
{\vcenter{\offinterlineskip\halign{\hfil$\textstyle##$\hfil\cr
<\cr\sim\cr}}}
{\vcenter{\offinterlineskip\halign{\hfil$\scriptstyle##$\hfil\cr
<\cr\sim\cr}}}
{\vcenter{\offinterlineskip\halign{\hfil$\scriptscriptstyle##$\hfil\cr
<\cr\sim\cr}}}}}

\newcommand{\re}[1]{{\cal R}{\rm e}[#1]}
\newcommand{\im}[1]{{\cal I}{\rm m}[#1]}
\newcommand{\farcs}{''}
\begin{document}
\title{Imaging binary stars by the cross-correlation technique\footnote{Based on observations made at 2\,m Télescope Bernard Lyot, Pic du Midi, France.}}
\author{\'E. Aristidi,  M. Carbillet, J.-F. Lyon, C. Aime\\ 
        U.M.R. 6525 Astrophysique Universit\'e de Nice --\\ Sophia Antipolis \\ Centre National de la Recherche Scientifique,
\\ Parc Valrose, 
 06108 Nice Cedex 2, France}
\date{Received; accepted}
\maketitle
\begin{abstract}
We present in this paper a technique for imaging binary stars from speckle data. This technique is based upon the computation of the cross-correlation between the speckle frames and their square. This may be considered as a simple, easy to implement, complementary computation to the autocorrelation function of Labeyrie's technique for a rapid determination of the position angle of binary systems.  Angular separation, absolute position angle and relative photometry of binary stars can be derived from this technique. We show an application to the bright double star $\zeta$\,Sge observed at the 2\,m Télescope Bernard~Lyot.  

\end{abstract}

\section{Introduction}
Processing binary stars by speckle interferometry (Labeyrie, 1970) leads to a 180$^\circ$ ambiguity in the measured position angle (PA). This is known as ``quadrant ambiguity''. Several techniques of speckle imaging can solve the problem, among which the techniques of Knox-Thompson (Knox and Thompson, 1974), shift-and-add (Bates, 1982) and speckle masking (Weigelt, 1991). A review of these techniques has been made by Roddier (Roddier, 1988). As they aim to reconstruct the image of any extended object from its specklegrams, these techniques usually require a lot of computer resources and processing time. They are not really well adapted to the double star problem: observers want to measure the separation and the PA of many stars a night and need a fast (near real-time) processing. Several techniques have been suggested for this purpose; for example the Directed Vector Autocorrelation (Bagnuolo {\em et al.}, 1992) which provides both the separation and absolute PA, the ``fork'' algorithm (Bagnuolo, 1988) based on the analysis of four equidistant points in the double star's specklegrams or the probability imaging technique (Carbillet, 1996b) based on the computation of twofold probability density functions of the specklegrams. These later techniques require a prior knowledge of the star separation which is usually measured from the power spectrum.

We propose a technique based upon the computation of a quantity very close to the autocorrelation function (AC): the cross-correlation (CC) between the specklegrams and their square. This function can be written as a slice of the triple correlation obtained for a {\em speckle masking vector} equal to zero. It is a two-dimensional function. For a double star, this quantity at first glance looks like the AC: a central peak surrounded by two smaller ones. These secondary peaks, identical in the AC, are asymmetric for the CC, allowing a quick diagnostic of the relative position of the two stars. The CC is almost as easy to compute as the AC, does not require the prior estimation of the power spectrum, and is then suitable for real-time processing. It also permits, under some hypothesis which will be developed in the text, the determination of the magnitude difference between the stars. 

This paper is organized as follows. Section 2.1 defines the statistical function we use, and derives relevant expressions for the double star. Section 2.2 describes the technique proposed to process real star data. We shall see in particular that the object-image convolution relation valid for the AC does not apply here and we propose a solution to overcome this difficulty. Section 3 is devoted to low-light level and photon bias. Application of the CC technique is investigated for clipped photon-counting specklegrams (where the number of detected photons is ``0'' or ``1'').

\begin{figure*}
\center\includegraphics{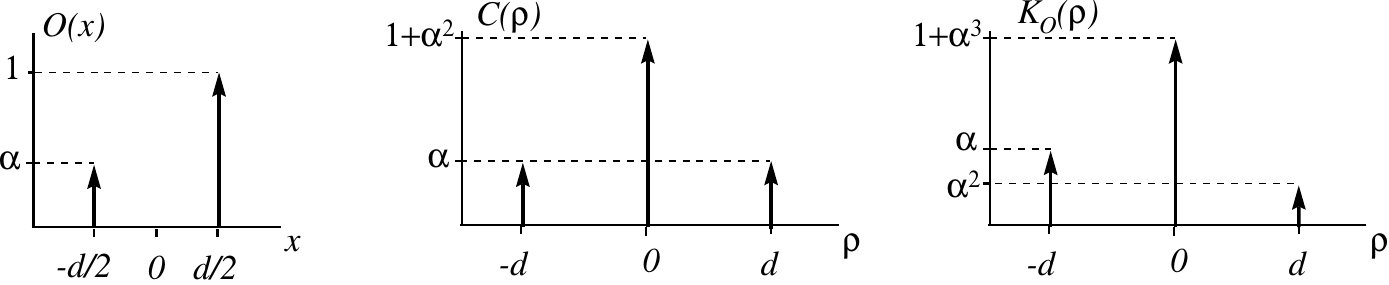}
\caption{Schematic representation of a double star $O(x)$ (left), its AC $C(\rho)$ (middle) and the CC $K_O(\rho)$ between $O(x)$ and its square (right). The arrows represent Dirac delta distributions. Note the asymmetry of the CC, where the ratio between the intensities of the two peaks in $(-d)$ and (+d) is exactly the intensity ratio of the stars.}\label{fig_ac_ic}
\end{figure*}

\section{General expressions}
\subsection{Cross-correlation/spectrum between a double star's image and its square}
In this paper one-dimensional notation will be used for simplicity, the extension to two dimensions being trivial.
The intensity of a double star $O(x)$ can be modeled as the sum of two unit impulses distant $d$ and weighted by the intensity ratio $\alpha$, i.e.:
\begin{equation}
O(x)=\delta(x-\frac{d}{2})+\alpha \delta(x+\frac{d}{2})
\end{equation}
\subsubsection*{\bf Cross-correlation}
We denote as $K_O(\rho)$ the cross-correlation (CC) between $O(x)$ and its square. $K_O(\rho)$ is defined as
\begin{equation}
K_O(\rho)=\int_{-\infty}^{\infty} O^2(x)\, O(x+\rho)\, dx
\label{eq_def_K}
\end{equation}
This function is a slice of the triple correlation of $O(x)$ defined as (Weigelt, 1991)
\begin{equation}
T_O(\rho_1,\rho_2)\; = \; \int_{-\infty}^{\infty}
O(x)\, O(x+\rho_1)\, O(x+\rho_2)\, dx 
\label{eq:eq2}
\end{equation}
we have $K_O(\rho)=T_O(0,\rho)$.\\
For a double star, $K_O(\rho)$ becomes
\begin{equation}
K_O(\rho)=(1+\alpha^3)\delta(\rho) \; + \; \alpha^2 \delta(\rho-d) \; + \; \alpha \delta(\rho+d)
\end{equation}
This quantity may be compared with the AC $C(\rho)$ of the double star $O(x)$
\begin{equation}
C(\rho)=(1+\alpha^2)\delta(\rho) \; + \; \alpha \delta(\rho-d) \; + \; \alpha \delta(\rho+d)
\end{equation}
Both $C(\rho)$ and $K_O(\rho)$ are composed of a central peak surrounded by two smaller ones distant $d$ (see figure~\ref{fig_ac_ic}). For the AC, these two peaks are symmetrical whatever the value of $\alpha$. This is why Labeyrie's speckle interferometry cannot give the relative positions of the two stars when observing a binary system. The CC $K_O(\rho)$ presents two asymmetrical peaks of ratio $\alpha$. The relative position of the peaks is those of the stars in $O(x)$. Using this quantity in double star's speckle interferometry, rather than AC, should give the position angle (PA) of the binary without any ambiguity.\\
\subsubsection*{\bf Cross-spectrum}
In the Fourier domain, the cross spectrum (CS) $\hat{K}_O(u)$ between $O(x)$ and its square is the Fourier transform of $K_O(\rho)$. It is a complex quantity whose real and imaginary parts are:\\

\begin{equation}
\begin{array}{l}
\displaystyle \re{\hat{K}_O(u)} \; = \; 1+\alpha^3 +
\alpha (1+\alpha) \cos(2\pi u d) \\ \\

\displaystyle \im{\hat{K}_O(u)} \; = \; \alpha (\alpha-1)\: \sin(2\pi u d)

\end{array}
\end{equation}\ \\

\begin{figure*}
\center\includegraphics{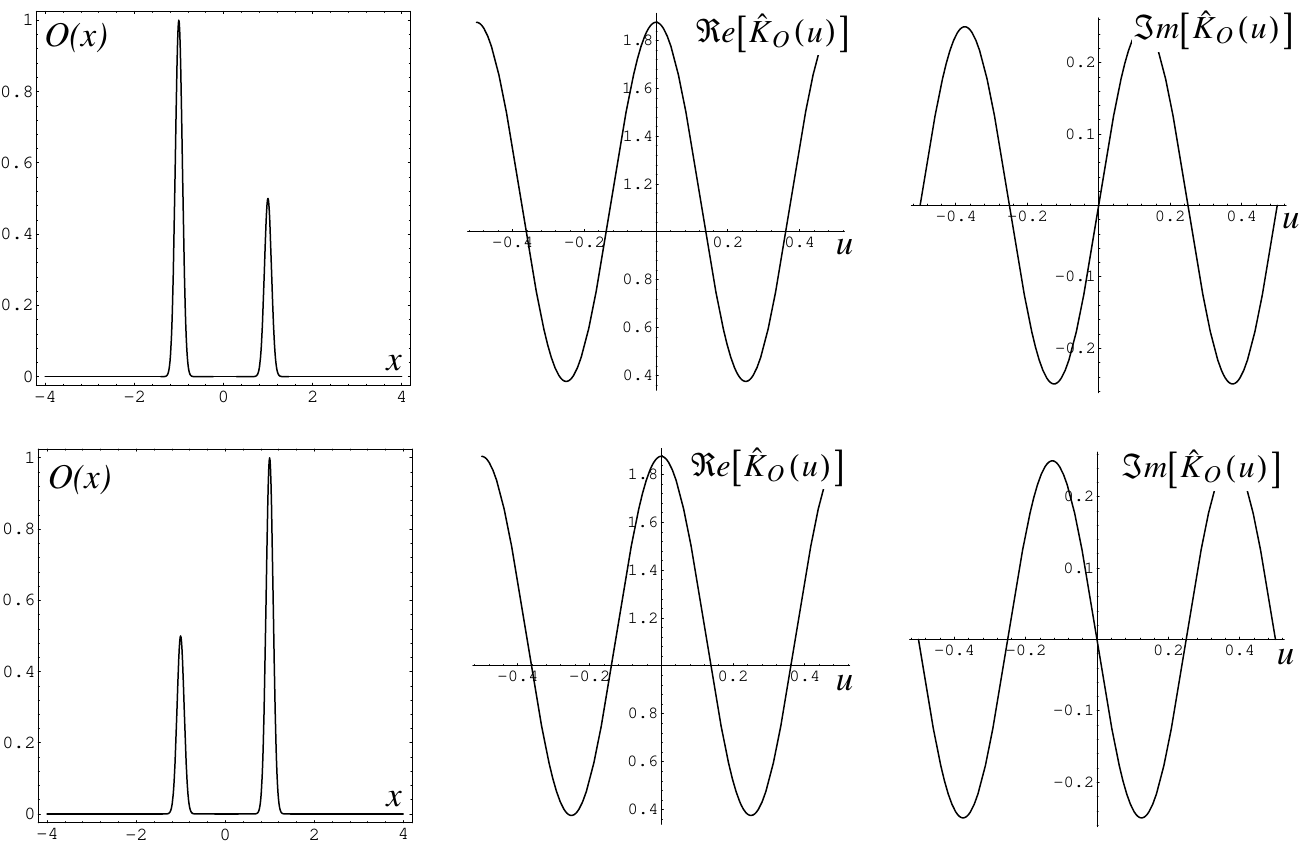}
\caption{Real and imaginary parts of the CS between a double star $O(x)$ and its square. Both figures are for an intensity ratio $\alpha=0.5$ between the two stars. Up: brighter star on the left, down: brighter star on the right. The real part of the CS is not sensitive to this orientation contrary to the imaginary part: its slope at the origin is positive in the first case and negative in the second one.
}
\label{fig_slope}
\end{figure*}

Both are sinusoidal functions of period $\frac{1}{d}$. The amplitude of the real and of the imaginary part gives the value of $\alpha$ without any ambiguity. But information concerning the relative position of the stars is fully contained in the imaginary part of $\hat{K}_O(u)$. Let $s$ be the slope of $\im{\hat{K}_O(u)}$ at the origin: 
\begin{equation}
s \; = \; \left[\frac{d}{du} \im{\hat{K}_O} \right]_{u=0} \; = \; 
2 \pi d \alpha(\alpha-1)
\label{eq:def_slope}
\end{equation}
We note that $s<0$ when $\alpha>1$ and $s\ge 0$ when $\alpha \le 1$. See figure~\ref{fig_slope} for illustration.

\subsection{Estimation of $\hat{K}_O(u)$ from speckle data}
We denote as $I(x)$ the instantaneous double star's specklegrams and $S(x)$ the corresponding point-spread function (PSF). Assuming isoplanatism, we can write\\

\begin{equation}
I(x)=S(x-\frac{d}{2}) + \alpha S(x+\frac{d}{2})
\label{eq:i(x)}
\end{equation}\ \\
We denote as $K_I(\rho)$ the CC of $I(x)$ and $K_O(\rho)$ the CC of $S(x)$.
\begin{equation}
K_I(\rho)=\langle \int_{-\infty}^{\infty} I^2(x)\, I(x+\rho)\, dx \rangle
\label{eq:ki(rho)}
\end{equation}
where $\langle \rangle$ denotes ensemble average. From eqs.~\ref{eq_def_K} and~\ref{eq:eq2}, we have
$K_I(\rho)=T_I(0,\rho)$ and $K_S(\rho)=T_S(0,\rho)$.

Unfortunately we can't find between $K_I(\rho)$ and $K_S(\rho)$ the simple convolution relation that exists between the corresponding full triple correlations. Inserting the value of $I(x)$ of eq.~\ref{eq:i(x)} into eq.~\ref{eq:ki(rho)}, a simple calculation gives:
\begin{equation}
\begin{array}{ll}
K_I(\rho) = & (1+\alpha^3) K_S(\rho) + \alpha K_S(\rho+d) + \alpha^2 K_S(\rho-d) \\ 
         &  +2 \alpha T_S(d,\rho)+2 \alpha^2 T_S(-d,\rho)
\end{array}
\end{equation}
This can be written as a convolution product plus a bias term
\begin{equation}
K_I(\rho)= K_S(\rho) \ast K_O(\rho) + B(\rho)
\label{eq:convplusbias}
\end{equation}
where the bias term is
\begin{equation}
 B(\rho)= 2 \alpha T_S(d,\rho)+2 \alpha^2 T_S(-d,\rho)
\end{equation}

It is difficult to estimate and subtract this bias from speckle data because of the presence of the unknown factors $2\alpha$ and $2\alpha^2$. Nevertheless we shall see that $B(\rho)$ vanishes if we consider zero-mean specklegrams in the case of a star separation large with respect to the speckle size $s$.\\

We call $\tilde{S}(x)$ and $\tilde{I}(x)$ the zero-mean specklegrams of the PSF and the double star:
\begin{equation}
\begin{array}{l}
\displaystyle \tilde{S}(x)=S(x)-\bar{S} \\ \\
\displaystyle \tilde{I}(x)=I(x)-\bar{I} \\ \\
\end{array}
\end{equation}
We respectively denote as $\tilde{m}_S$, $C_{\tilde{S}}(\rho)$, $K_{\tilde{S}}(\rho)$ and $T_{\tilde{S}}(\rho_1,\rho_2)$ the mean (with obviously $\tilde{m}_S=0$), the AC, the CC and the triple correlation of $\tilde{S}(x)$. We denote as $K_{\tilde{I}}(\rho)$ the CC of $\tilde{I}(x)$. From eqs.~12--14 we have
\begin{equation}
K_{\tilde{I}}(\rho) = K_{\tilde{S}}(\rho) \ast K_{O}(\rho) + 2\alpha T_{\tilde{S}}(d,\rho) + 2\alpha^2 T_{\tilde{S}}(-d,\rho)
\end{equation}
Let us consider the term $T_{\tilde{S}}(d,\rho)$. We have
\begin{equation}
\begin{array}{ll}
T_{\tilde{S}}(d,\rho) & = \langle \int \tilde{S}(x) \tilde{S}(x+d) \tilde{S}(x+\rho) dx \rangle \\
     & = E[\tilde{S}(x) \tilde{S}(x+d) \tilde{S}(x+\rho)]
\end{array}
\end{equation}
where $E[\bullet]$ is the mathematical expectation of $\bullet$. We have assumed that $d\gg s$, so $\tilde{S}(x)$ and $\tilde{S}(x+d)$ are uncorrelated. We can distinguish 3 cases:
\begin{enumerate}
\item $\rho\lsim s$:\\
$\tilde{S}(x)$ and $\tilde{S}(x+\rho)$ are correlated; $\tilde{S}(x+d)$ is uncorrelated both with $\tilde{S}(x)$ and $\tilde{S}(x+\rho)$, so:
\begin{equation}
\begin{array}{ll}
T_{\tilde{S}}(d,\rho) & = E[\tilde{S}(x+d)].E[\tilde{S}(x) \tilde{S}(x+\rho)] \\
                      & = \tilde{m}_S C_{\tilde{S}}(\rho) = 0
\end{array}
\end{equation}

\item $|\rho-d|\lsim s$:\\
$\tilde{S}(x+d)$ and $\tilde{S}(x+\rho)$ are correlated; $\tilde{S}(x)$ is uncorrelated with the two others, so:
\begin{equation}
\begin{array}{ll}
T_{\tilde{S}}(d,\rho) & = E[\tilde{S}(x)].E[\tilde{S}(x+d) \tilde{S}(x+\rho)]\\
                      & = \tilde{m}_S C_{\tilde{S}}(\rho-d) = 0
\end{array}
\end{equation}

\item Otherwise:\\
Both $\tilde{S}(x)$, $\tilde{S}(x+d)$ and $\tilde{S}(x+\rho)$ are uncorrelated, so:
\begin{equation}
\begin{array}{ll}
T_{\tilde{S}}(d,\rho) & = E[\tilde{S}(x)].E[\tilde{S}(x+d)].E[\tilde{S}(x+\rho)]\\
                      & = \tilde{m}_S^3 = 0
\end{array}
\end{equation}
\end{enumerate}
\begin{figure*}
\includegraphics{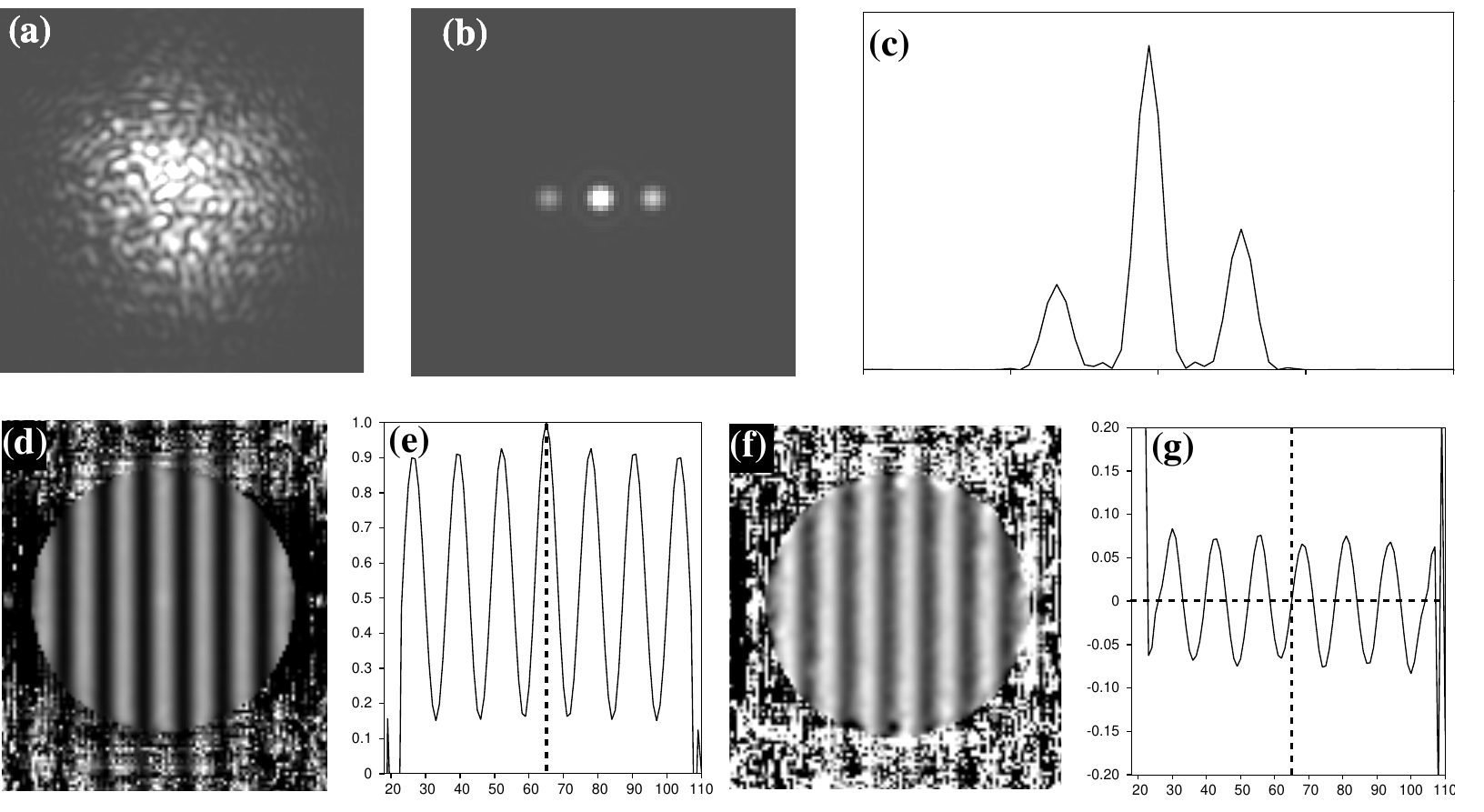}
\caption{Cross-correlation/spectrum computed on simulated speckle patterns. The calculus has been made on two sets of 200 images, one for the double star and one for the reference star. The double star is 10~pixels separation oriented along the $x$-axis. The intensity ratio  is 0.5. The simulation has been made for a Fried parameter $r_0=20$~cm, a telescope diameter of 2.60~m and a wavelength $\lambda=500$~nm. Picture (a) is a typical double star's specklegram, (b) is the two-dimensional object's CC $K_O(\rho)$ and (c) is a cut along the $\rho_x$ axis. Notice the asymmetry of the two secondary peaks. Lower pictures are the real (d) and imaginary (f) parts of $\hat{K}_O(u)$, while the curves (e) and (g) are the corresponding cuts along the $u_x$ axis. Note the sign of the slope at the origin of $\im{\hat{K}_O(u)}$.}
\label{fig_simu}
\end{figure*}

\begin{figure*}
\includegraphics{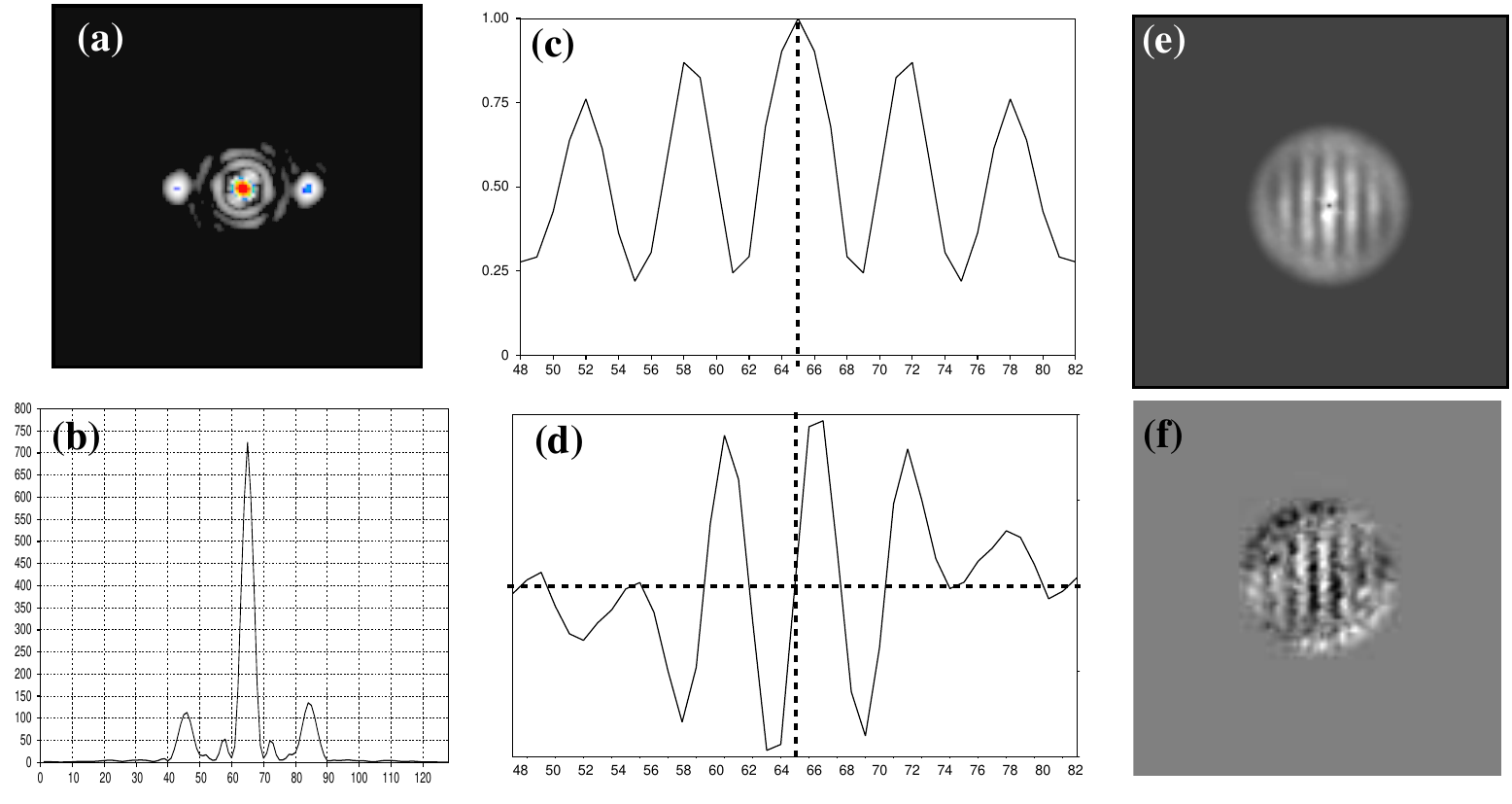}
\caption{This figure shows an application to the bright star $\zeta~Sge$ (see text for details). Computation was made for 1089 short exposure (20 ms) frames of the double star and 2993 on the reference star HR\,7536. The mean value of each specklegram has been estimated as an average of the intensity over the image, then subtracted. Figures (a) and (b) are the two-dimensional CC and its cut along the $\rho_x$ axis (the coordinate system has been rotated so that interesting features are along the horizontal axis). The asymmetry of the secondary peaks gives the relative position of the stars. Figures (e) and (f) are the real and imaginary parts of the CS $\hat{K}_O(u)$, (c) and (d) are cuts along the $u_x$ axis. The imaginary part of $\hat{K}_O(u)$ is a sine function with positive slope at origin: the brightest star is on the left. }
\label{fig_betadel}
\end{figure*}
The term $T_{\tilde{S}}(-d,\rho)$ is obtained by changing $d$ into $-d$ in the above expressions. We see that in most cases the bias vanishes under the hypothesis $d \gg s$. It is important to remark that this previous calculus is valid only under the space-stationarity hypothesis, i.e. if $m_S$ is the same on the whole image. This is valid only if we take the central part of the speckle pattern.\\
Let us assume that $d \gg s$. We can then write the approximation
\begin{equation}
K_{\tilde{I}}(\rho) = K_{\tilde{S}}(\rho) \ast K_{O}(\rho)
\label{eq_est_ko}
\end{equation}
and in the Fourier domain:
\begin{equation}
\hat{K}_{\tilde{I}}(u) = \hat{K}_{\tilde{S}}(u) . \hat{K}_{O}(u)
\label{eq_est_ku}
\end{equation}
Estimating $\hat{K}_O(u)$ from speckle data is very similar to classical speckle interferometry processing. The cross-spectra are estimated as ensemble averages (${\cal F}$ denoting the Fourier Transform):
\begin{equation}
\begin{array}{l}
\displaystyle \hat{K}_{\tilde{I}} =\langle {\cal F}[{\tilde{I}^2}] {\cal F}[{\tilde{I}}]^*\rangle\\ \\
\displaystyle \hat{K}_{\tilde{S}} =\langle {\cal F}[{\tilde{S}^2}] {\cal F}[{\tilde{S}}]^*\rangle
\end{array}
\end{equation}
Note that $\hat{K}_S(u)$ is a real function (assuming the statistical properties of the ideal point-spread speckle pattern are isotropic in space).
 
Numerical simulations of speckle data are presented in figure~\ref{fig_simu}.
This technique has been applied successfully to the newly discovered double star {\sc Moai}~1 (Carbillet {\em et al.}, 1996a). Figure~\ref{fig_betadel} shows another application to the star $\zeta$~Sge. Observations were made on September, 1994 with the 2m Télescope Bernard Lyot (TBL) of the Pic du Midi observatory, using the speckle camera of the Aperture Synthesis group of Observatoire Midi-Pyr\'en\'ees (Andr\'e {\em el al.}, 1994) and an ICCD detector.

\section{Low light level}
\subsection{Expression of the photon bias in the cross-correlation}
In this subsection we denote as the generic name $K(\rho)$ one of the functions $K_O(\rho)$, $K_{I}(\rho)$ or $K_{S}(\rho)$. The same is for their Fourier transforms: $\hat{K}(u)$. These functions are the CC and the CS of a high-light level zero-mean speckle pattern.

Since $K(\rho)$ is a slice of the triple correlation of $O(x)$, it is possible to take advantage of the calculus of the bias terms made by Aime {\em et al.} (1992) in the photodetected triple correlation. Equation~2.18 of this last paper leads to the following expression for the photodetected cross-correlation $K_p(\rho)$ of the zero-mean
\begin{equation}
K_p(\rho) \; = \; \bar{N}^3 K(\rho) +  B_p(\rho)
\end{equation}
$B_p(\rho)$ is a photon bias term whose expression is
\begin{equation}
B_p(\rho) \; = \; 2 \bar{N}^2 C(0) \delta(\rho) + \bar{N}^2 C(\rho) + \bar{N} \delta(\rho)+\bar{N}
\end{equation}
where $\bar{N}$ is the average number of photons per image, $C(\rho)$ is the correlation function of the zero-mean high-light level speckle pattern (standing for $C_O(\rho)$, $C_{I}(\rho)$ and $C_{S}(\rho)$) and $m$ is its mean. The bias terms are not as simple as for the photodetected AC (Aime {\em et al.}, 1992) where it is just a Dirac delta function at the origin.
The photodetected CS $\hat{K}_p(u)$ is biased by frequency-dependent terms
\begin{equation}
\hat{K}_p(u) \; = \;  2 \bar{N}^2 C(0) + \bar{N}^2 W(u) + \bar{N} + \bar{N}^3 \hat{K}(u)
\label{eq:isphot}
\end{equation}
where $W(u)$ is the power spectrum, Fourier transform of $C(\rho)$.  It is remarkable to notice that bias terms are real. The imaginary part of the photodetected cross-spectrum is unbiased. Its expression is 
\begin{equation}
\im{\hat{K}_p(u)} \; = \; \bar{N}^3 \im{\hat{K}(u)}
\label{eq_kp_im}
\end{equation}
\subsection{Case of a bright reference star}
For a bright enough reference star, the detection at high light level of the specklegrams $S(x)$ allows to compute the high-light level zero-mean cross-spectrum $\hat{K}_{\tilde{S}}(u)$. We assume that the specklegrams $I(x)$ are detected in photon-counting mode. We denote as $\hat{K}_{Op}(u)$ the ratio between the photodetected cross-spectrum of $I(x)$ and the cross-spectrum of $S(x)$
\begin{equation}
\hat{K}_{Op}(u) \; = \; \frac{\hat{K}_{Ip}(u)}{\hat{K}_{\tilde{S}}(u)}
\end{equation}
Even in the case of a well resolved double star where the convolution relation may be applied, 
$\hat{K}_{Op}(u)$ is not a good estimator of the double star cross-spectrum because of the complicated bias terms. Its real and imaginary parts are
\begin{equation}
\begin{array}{ll}
\re{\hat{K}_{Op}(u)} = &  \bar{N}_I^3 \re{\hat{K}_O(u)} + \frac{1}{\hat{K}_{\tilde{S}}(u)} (2 \bar{N}_I^2 C_{I}(0)\\
                       & + \bar{N}_I^2 W_{I}(u) + \bar{N}_I)
\\ \\
\im{\hat{K}_{Op}(u)} = & \bar{N}_I^3 \im{\hat{K}_O(u)}
\end{array}
\end{equation}
where $N_I$ is the average number of photons per image in the specklegrams of $I(x)$. Here again it appears that the imaginary part of $\hat{K}_{Op}(u)$ is unbiased. This may be interesting if we remember that this imaginary part contains the information on the relative position of the stars in $O(x)$.

\begin{figure*}
\includegraphics{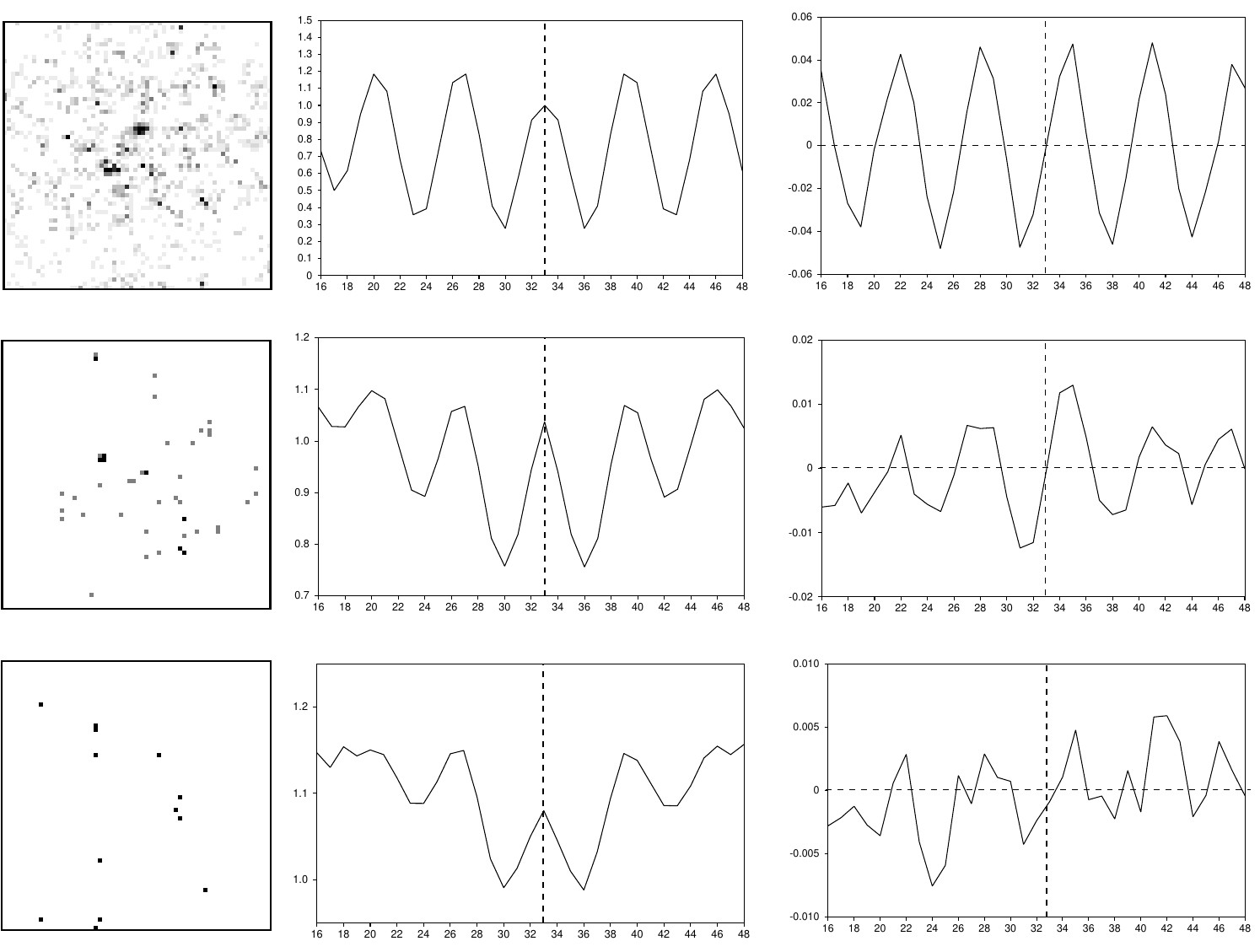}
\caption{Simulation of photodetected CS for different number of photons per image. The upper figures are for 2000 photons/frame, the middle are for 40 photons/frame and the lower ones are for 15 photons/frame. The computation was made on two sets of 5000 images with the same parameters than figure 3. The pictures on the left are typical specklegrams. Curves are real (middle) and imaginary (right) parts of the biased object's cross spectrum $\hat{K}_{Op}(u)$ estimated as indicated in the text. Notice that even at the lowest light level (15 photons/frame) it is possible to predict the relative position of the stars using the sign of the slope at the origin $s_p$ of the imaginary part.}
\label{fig-simu-ph}
\end{figure*}

\subsection{General case}
In this subsection we suppose that both $I(x)$ and $S(x)$ are photodetected. We denote as $N_S$ the average number of photons per image in the specklegrams of $S(x)$. We shall see that the information on the relative position of the stars is still present. This information is contained in the slope of the imaginary part of $\hat{K}_O(u)$ at the origin (see figure~\ref{fig_slope}). For a high-light level detection where $\hat{K}_O(u)$ is estimated as written in equation~\ref{eq_est_ku}, the slope $s$, defined in eq.~\ref{eq:def_slope}, can be written as (after a few algebra)
\begin{equation}
s\; =\left[\frac{d}{du} \im{\hat{K}_O} \right]_{u=0} \; = \; \frac{1}{\hat{K}_{S}(0)} \left[\frac{d}{du} \im{\hat{K}_{I}} \right]_{u=0}
\end{equation}
where we use the fact that $\im{\hat{K}_{I}(0)}=0$. The sign of $s$ is that of $\left[\frac{d}{du} \im{\hat{K}_{I}} \right]_{u=0}$.\\

We denote as $s_p$ the slope at the origin of $\hat{K}_{Op}(u)$
defined as the ratio between the photodetected cross-spectra of $I(x)$ and $S(x)$. The expression of $s_p$ is similar to the previous equation 
\begin{equation}
s_p\; = \; \frac{1}{\hat{K}_{Sp}(0)} \left[\frac{d}{du} \im{\hat{K}_{Ip}} \right]_{u=0}
\end{equation}
and from equation~\ref{eq_kp_im} $s_p$ expresses as
\begin{equation}
s_p\; = \; \frac{\bar{N}_I^3}{\hat{K}_{Sp}(0)} \left[\frac{d}{du} \im{\hat{K}_{I}} \right]_{u=0}
\end{equation}
Taking the expressions given in equation~\ref{eq:isphot} for $\hat{K}_{Sp}(0)$,
\begin{equation}
s_p\; = \; \frac{\bar{N}_I^3 \left[\frac{d}{du} \im{\hat{K}_{I}} \right]_{u=0}}
                {2 \bar{N}_S C_{S}(0) +  \bar{N}_S^2 W_{S}(0) +  \bar{N}_S^3 \hat{K}_{S}(0)}
\end{equation}
in the case of $N_S\gg 1$ this relation may be approximated by
\begin{equation}
s_p = \frac{\bar{N}_I^3}{\bar{N}_S^3} s
\end{equation}

\begin{figure*}
\includegraphics{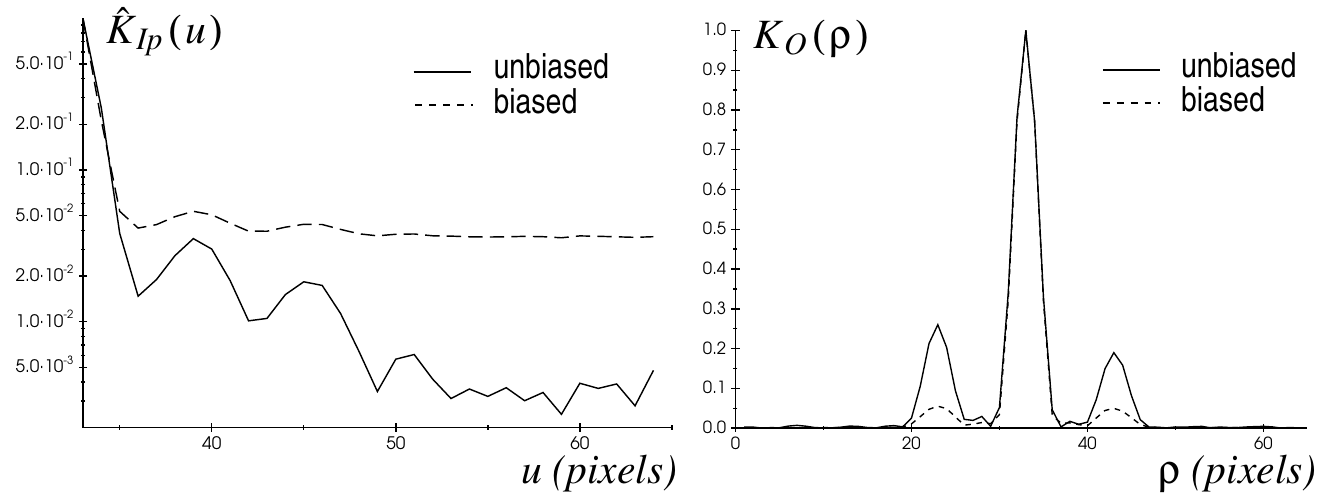}
\caption{Simulation of 10000 photodetected images of a double star of separation $d=10$~pixels and intensity ratio $\alpha=0.5$. The average number of photons per image is 50. The parameters of the simulation are 2.60~m for the telescope diameter, $r_0=30$~cm and wavelength $\lambda=5000$~\AA. 10000 images of a single reference star have also been synthesized with the same conditions. Curves on the left are the real part of the photon-biased and unbiased CS $\hat{K}_{Ip}(u)$ of the double star's images. Curves on the right are the biased and unbiased object's CC $K_O(\rho)$. The ratio of the two peaks is 0.90 for the biased data and 0.73 for the unbiased ones.
}
\label{fig:debiais}
\end{figure*} 

The signs of $s$ and $s_p$ are the same because the denominator of eq.~31 is positive. The relative position of the stars can then be retrieved in spite of the photon bias. Results of a simulation are shown in figure~\ref{fig-simu-ph}.

\subsection{Subtracting the photon bias}
The frequency-dependent bias terms in the expression of $\hat{K}_p(u)$ can easily be removed by subtracting the photodetected power spectrum $W_p(u)$, whose expression is derived from Aime {\em et al.} (1992) and is valid in the case where the high-light level mean is zero
\begin{equation}
W_p(u)= \bar{N} ^2 W(u) + \bar{N}
\end{equation}
From equation~\ref{eq:isphot} it appears that 
\begin{equation}
\hat{K}_p(u)=\bar{N}^3 K(u) + W_p(u) + 2 \bar{N}^2 C(0)
\end{equation}
This bias is quite easy to remove when processing real data. $\hat{K}_p(u)$ and $W_p(u)$ are computed directly from the data, then subtracted. The remaining bias is the constant $2 \bar{N}^2 C(0)$ and can be estimated beyond the cutoff frequency.

The efficiency of this bias subtraction is shown in figure~6. It is a simulation of 10000 photon-counting specklegrams (50 photons/image) of a double star  with a separation of 10~pixels and an intensity ratio of 0.5. As expected, the major improvement of the bias subtraction is to restore the asymmetry of the cross-correlation's secondary two peaks, thus allowing a better diagnostic of the relative position of the two stars.

\subsection{Clipping conditions}
Some photon-counting detectors have centroiding electronics which compute in real time the photon coordinates and cannot distinguish between one photon and more photons which have come onto a given pixel during the integration time. Intensities on the specklegrams are then thresholded to ``1'' and this is what we call ``clipping''. Such images are then equal to their square and the CC is equal to the AC. The asymmetry is lost.\\
We propose computing alternative quantities to solve this problem. The first one is:
\begin{equation}
\Xi_1(\rho) \; = \; \langle \int I(x) I(x+\epsilon) I(x+\rho) \, dx\rangle
\end{equation}
If $\epsilon$ is small compared to the speckle size, but larger than the ``centreur hole'' size (Foy, 1987), $I(x)$ and $I(x+\epsilon)$ are correlated enough to provide $\Xi_1(\rho)$ with the same properties than the CC.\\
The following function may also be computed but it requires the knowledge of the star separation $d$:
\begin{equation}
\Xi_2(\rho) \; = \; \langle \int I(x) I(x+d) I(x+\rho) \, dx\rangle
\end{equation}

\begin{figure*}
\includegraphics{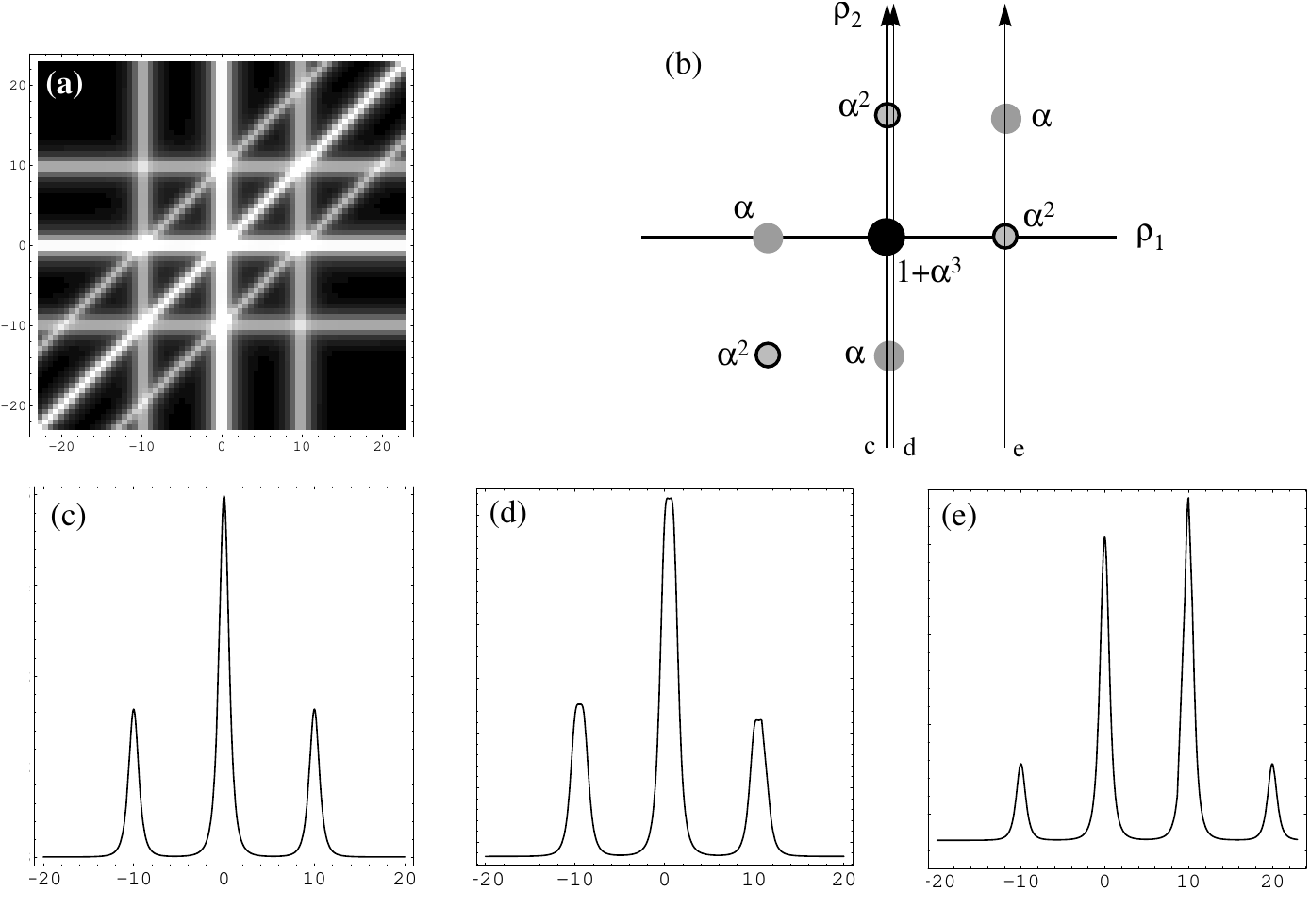}
\caption{Picture (a) is the triple correlation $T(\rho_1,\rho_2)$ of clipped photon counting specklegrams of a double star. It is a plot in the $(\rho_1,\rho_2)$ plane in the case where $\rho_1$, $\rho_2$ and $d$ are collinear. The double star has a separation of 10~pixels and an intensity ratio $\alpha=0.5$. (b) is a schematic representation of (a) where the relevant peaks are drawn as filled circles with values of the TC indicated. Curves (c), (d) and (e) are the CC, the function $\Xi_1(\rho)$ and the function $\Xi_2(\rho)$. These curves correspond to slices of the TC along the directions indicated by the vertical arrows. As expected the CC does not show any asymmetry. The function $\Xi_1(\rho)$ looks similar to the unclipped CC with a slight asymmetry between the peaks (no photon bias correction has been applied here). The function $\Xi_2(\rho)$ presents four peaks. The two external ones are due to photon bias. The two central ones contain information about the relative position . Their asymmetry is in the opposite sense than those of the secondary peaks of the CC.
}
\label{fig:tcclip}
\end{figure*}
\begin{figure*}
\includegraphics{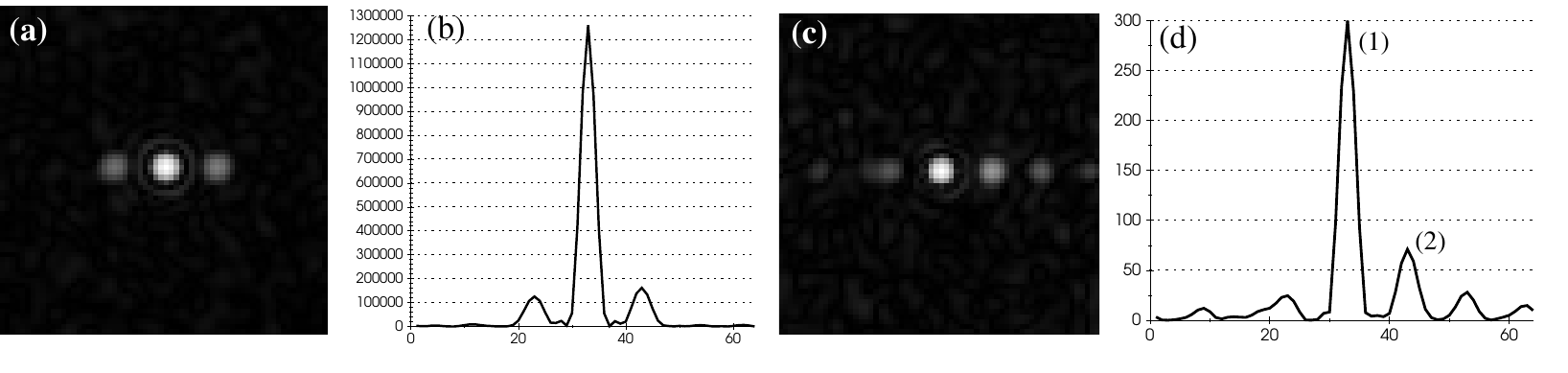}
\caption{Simulation of clipped photon-counting specklegrams. 1000 frames have been simulated for a double star of separation 10 pixels and intensity ratio 0.5. 1000 frames of a point source have also been simulated. Parameters of the simulation are: telescope diameter: 2.60\,m, Fried parameter: 30\,cm, wavelength: 500~nm and number of photons per frame: 200 (each frame is $64\times 64$ pixels). Picture (a) is the function  $\Xi_1(\rho)$ of the double star divided by those of the reference star. Curve (b) is a slice along the $\rho_x$ axis. Figures (c) and (d) are the same for the function $\Xi_2(\rho)$. The two peaks (1) and (2) give the information about the relative position of the stars (brighter star on the right in this simulation). The other peaks of $\Xi_2$ are ghosts due to photon bias.
}
\label{fig:ksiclip}
\end{figure*}
\pagebreak[1]

\begin{figure*}[t]
\includegraphics{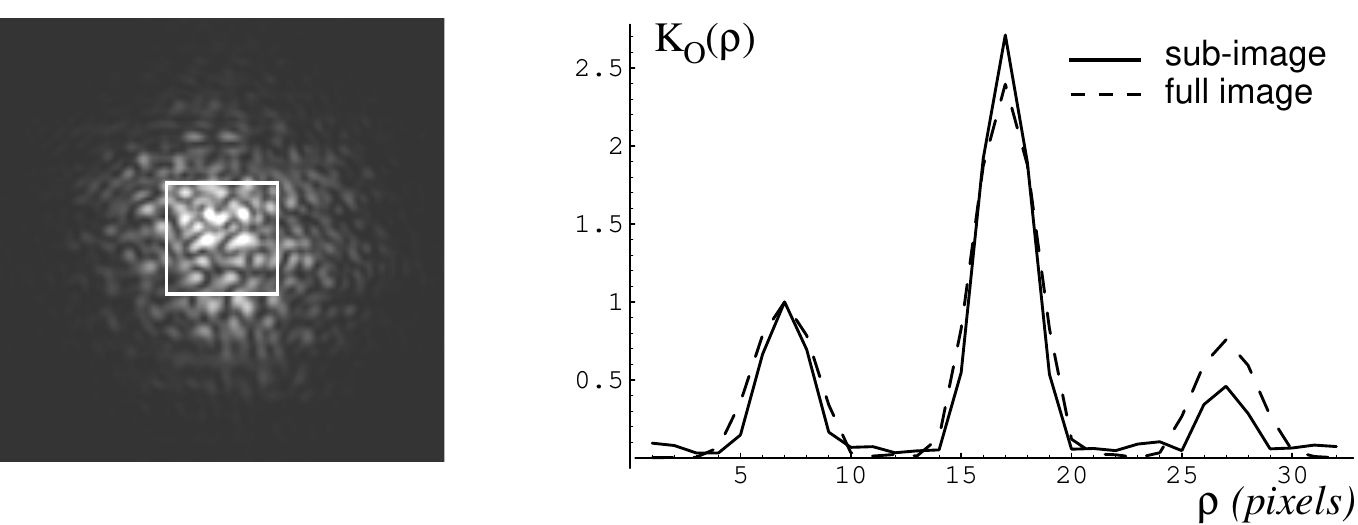}
\caption{Simulation of 100 specklegrams of a double star with the same parameters than figure~3. The image size is 128x128. Left: a typical specklegram. the white square on the left figure demarcates a sub-image of 32x32 centered on the photocenter of the speckle pattern. The full-width-middle height of the speckle pattern shape is $\simeq 34$ pixels. Right: a cut of the object's CC $K_O(\rho)$ after deconvolution by a reference star's CC computed from 100 specklegrams. Dashed  line: computation on the whole images, full  line: computation on the sub-images. In both cases, the mean of the specklegrams has been estimated as the average of the intensity on the image, then subtracted. The CC computed on the whole images gives an intensity ratio $\alpha=0.76$. For the sub-images, the statistical mean is almost constant: the CC is almost unbiased and the ratio of the two secondary peaks gives $\alpha=0.46$ (actual value is 0.5).}
\label{fig:zeromeansubimages}
\end{figure*}

These two functions correspond to slices of the triple correlation: $\Xi_1(\rho)=T(\epsilon,\rho)$ and $\Xi_2(\rho)=T(d,\rho)$. In order to understand the behavior of these quantities, we have computed the triple correlation $T(\rho_1,\rho_2)$ of clipped photon-counting specklegrams of a double star, for fully-developed speckle patterns. In that case  the complex amplitude at the focal plane is a Gaussian random variable and analytical expressions  can be obtained for the clipped TC (Aristidi {\em et al.}, 1995). Figure~\ref{fig:tcclip} shows the TC and the functions $\Xi_1(\rho)$ and $\Xi_2(\rho)$ for the Gaussian hypothesis. The function $\Xi_1(\rho)$ has the same behavior than the unclipped CC: two asymmetrical peaks giving the couple orientation. The function $\Xi_2(\rho)$ is a bit more complicated. It should present two asymmetrical peaks separated $d$ (asymmetry is the opposite of those of the CC) but there are also two ``ghosts'' at spatial lags $\pm 2d$ caused by photon bias. Simulations have been performed on clipped photon-counting specklegrams. The results, presented in figure~\ref{fig:ksiclip} agree with the analytical model.

\section{Discussion}
The technique we propose here may be seen as a complement to Labeyrie's speckle interferometry for binary stars. The CC is as easy to interpret as the classical AC but provides the absolute PA of the stars as well. The CC is very easy to implement. It has the advantage to give a very simple result in the form of a direct 2D image so that it appears worth it to try that method for a quick analysis of the PA when doing double star observations.
The use of a reference star may not be necessary for position measurements. The secondary peaks and their asymmetry are usually easy to see on the double star's CC. In the Fourier plane, the imaginary part of the CS also reveals  the position of the brighter star by its slope at the origin. However, a reference star can enhance the asymmetry  of the CC for difficult objects (very small or very large magnitude difference).

For relative photometry measurements (the intensity ratio of the couple) a reference star must be used. A careful attention must then be given to the bias subtraction. As shown in section~2, the convolution relation between the double star's CC and the PSF's CC applies only for zero-mean specklegrams under space-stationarity hypothesis. If one of these conditions is not fulfilled, the deconvolution will give a biased result (the intensity ratio is estimated by the ratio of the heights of the two peaks). Space-stationarity is generally a wrong assumption for real specklegrams: they present a finite spatial extent depending upon seeing conditions. The statistical mean of the speckle patterns is then a function of the position and cannot be estimated by averaging the intensity over the whole images, as it is done usually. Obtaining zero-mean specklegrams in these conditions is not simple. For small separations, it can be useful to process small sub-images extracted around the photocenter of the specklegrams. If the dimension of these sub-images is small enough compared to the size of the speckle patterns, the statistical mean can be considered as nearly constant. It can then be estimated as the spatial mean of the intensity over the sub-images and subtracted. The smaller the sub-images are, the better it will work. A simulation is shown in figure~\ref{fig:zeromeansubimages}. This is not suitable for large separations. Various algorithms may be tried in that case. For example subtracting to each specklegram the corresponding long-exposure image averaged over some hundreds of instantaneous frames. Or fitting each specklegram by a smooth function like a Gaussian, then subtracting it. Actually this will increase in the processing the weight of the small values of the border of the image, and consequently the noise.

At low light level, the frequency-dependent photon bias can be removed by subtracting the power spectrum to the CS. Here again, this operation is not really necessary for position measurements: the relevant information is contained in the slope at the origin of the unbiased imaginary part of the double star's CS. But it considerably enhances the asymmetry of the two secondary peaks of the CC (as shown by figure~\ref{fig:debiais}).

This technique has been  successfully used over about 20 double stars observed at the Télescope Bernard Lyot between 1994 and 1995. All the measured PA were compatible with the orbit of the stars. These results have been submitted to {\em Astronomy and Astrophysics}.
During our last observing run, we discovered a 0\farcs 1-separated binary star ({\sc Moai 1}) with almost zero magnitude difference. Its CC was slightly asymmetric and we predicted a PA for this couple (Carbillet {\em et al.}, 1996a).
\section*{acknowledgements}
The authors whish to thank J.-L. Prieur (Observatoire Midi-Pyr\'en\'ees) for the use of his speckle camera and his cooperation during the observations

\end{document}